\documentclass[12pt]{iopart}

\usepackage{hyperref}
\hypersetup{colorlinks,linkcolor=red,citecolor=blue,urlcolor=blue}

\usepackage{graphicx}

\usepackage[abbr]{harvard}

\begin{document}

\title[]{Atomic and Molecular Data for Optical Stellar Spectroscopy}

\author{U.~Heiter$^1$, K.~Lind$^1$, M.~Asplund$^2$, P.~S.~Barklem$^1$, M.~Bergemann$^3$, L.~Magrini$^4$, T.~Masseron$^5$, \v{S}.~Mikolaitis$^{6,7}$, J.~C.~Pickering$^8$, M.~P.~Ruffoni$^8$}

\address{$^1$Institutionen f\"or fysik och astronomi, Uppsala universitet, Box 516, 751 20 Uppsala, Sweden}
\address{$^2$Research School of Astronomy and Astrophysics, Australian National University, Cotter Road, Weston Creek, ACT 2611, Australia}
\address{$^3$Max Planck Institute for Astronomy, K\"onigstuhl 17, 69117 Heidelberg, Germany}
\address{$^4$INAF - Osservatorio Astrofisico di Arcetri, Largo E. Fermi 5, 50125, Florence, Italy}
\address{$^5$Institute of Astronomy, University of Cambridge, Madingley Road, Cambridge CB3 0HA, United Kingdom}
\address{$^6$Laboratoire Lagrange (UMR7293), Universit\'e de Nice Sophia Antipolis, CNRS, Observatoire de la C\^ote d’Azur, BP 4229, 06304 Nice Cedex 04, France}
\address{$^7$Institute of Theoretical Physics and Astronomy, Vilnius University, A. Go\v{s}tauto 12, 01108 Vilnius, Lithuania}
\address{$^8$Blackett Laboratory, Imperial College London, London SW7 2BW, United Kingdom}
\ead{ulrike.heiter@physics.uu.se}
\vspace{10pt}
\begin{indented}
\item[] Received 19 Oct 2014 / Accepted 29 Dec 2014 by Phys. Scr.
\end{indented}

\begin{abstract}
High-precision spectroscopy of large stellar samples plays a crucial role for several topical issues in astrophysics. Examples include studying the chemical structure and evolution of the Milky Way galaxy, tracing the origin of chemical elements, and characterizing planetary host stars.
Data are accumulating from instruments that obtain high-quality spectra of stars in the ultraviolet, optical and infrared wavelength regions on a routine basis.
These instruments are located at ground-based 2- to 10-m class telescopes around the world, in addition to the spectrographs with unique capabilities available at the Hubble Space Telescope.
The interpretation of these spectra requires high-quality transition data for numerous species, in particular neutral and singly ionized atoms, and di- or triatomic molecules. We rely heavily on the continuous efforts of laboratory astrophysics groups that produce and improve the relevant experimental and theoretical atomic and molecular data.
The compilation of the best available data is facilitated by databases and electronic infrastructures such as the NIST Atomic Spectra Database, the VALD database, or the Virtual Atomic and Molecular Data Centre (VAMDC).
We illustrate the current status of atomic data for optical stellar spectra with the example of the Gaia-ESO Public Spectroscopic Survey.
Data sources for 35 chemical elements were reviewed in an effort to construct a line list for a homogeneous abundance analysis of up to $10^5$ stars.
\end{abstract}


\vspace{2pc}


%
%

\section{Stellar spectroscopy -- topical issues and data needs}

High-precision analysis of spectra of large numbers of stars is currently routinely used for
  \begin{itemize}
     \item studying the chemo-dynamical structure and evolution of the Milky Way galaxy,
     \item studying the evolution of stars,
     \item tracing the origin of chemical elements,
     \item or characterizing planetary host stars.
  \end{itemize}
High-quality spectra are accumulating in the archives of various observatories from dedicated surveys and individual programs.
The interpretation of these spectra requires high-quality data for numerous atomic and molecular transitions.

Large amounts of data are needed from the ultraviolet through the optical and into the infrared wavelength regions.
Stellar spectroscopy in the ultraviolet region is mainly based on high-resolution spectra obtained with current instruments on the Hubble Space Telescope.
The optical and infrared regions are surveyed with ground-based 2- to 10-m telescopes around the world. A few examples are the optical NARVAL spectrograph at the 2m telescope on Pic du Midi (France); the APOGEE (Apache Point Observatory Galactic Evolution Experiment) spectrograph at the 2.5m telescope in New Mexico (USA); the ESPaDOnS (Echelle SpectroPolarimetric Device for the Observation of Stars) spectrograph at the 3.6m telescope on Hawaii; the HARPS (High Accuracy Radial velocity Planet Searcher) spectrograph at the 3.6m telescope at ESO (Chile); and the UVES (Ultraviolet and Visual Echelle Spectrograph) and CRIRES (CRyogenic high-resolution InfraRed Echelle Spectrograph) instruments at the 8m VLT at ESO (Chile).

Cool stars with surface temperatures between about 3500~K and 6500~K are the most suitable objects to study the astrophysical topics mentioned above. Their spectra are dominated by absorption lines of mainly neutral and singly ionized atoms, as well as diatomic and triatomic molecules.

The types of data which are most important for spectrum analysis (apart from transition wavelengths, and assuming local thermodynamic equilibrium) are on the one hand transition probabilities (oscillator strengths, $gf$-values), and on the other hand data describing the effects of elastic collisions.
Oscillator strengths can be either measured by laboratory astrophysics groups or calculated by atomic physics groups.
Parameters for line-broadening by collisions with neutral or charged particles are obtained from calculations by atomic physics groups, e.g. \citeasnoun[and references therein]{2000A&AS..142..467B} 
or \citeasnoun[and references therein]{2014AdSpR..54.1148S}. 
Experimental data for collisional line broadening are only available for very few lines and have rather large uncertainties
\citeaffixed{1985JPhB...18...95L}{e.g.}.

%

Extraction of data from publications in regular scientific journals can be a tedious task. 
However, databases and electronic infrastructures are available to facilitate this part of the work of stellar spectroscopists. 
Examples for relevant databases are the NIST Atomic Spectra Database \cite{NIST}, the VALD database \cite{Pisk:95,Kupk:99,Ryab:99,2008JPhCS.130a2011H}\footnote{\url{http://vald.astro.uu.se}}, or the STARK-B database \cite{starkb}.
These databases can be queried either through their proprietory web interfaces, or through the VAMDC electronic infrastructure \cite{2010JQSRT.111.2151D}\footnote{Virtual Atomic and Molecular Data Centre, \url{http://www.vamdc.eu}}, 
which currently provides access to about 30 databases simultaneously.

\begin{figure}[ht]
   \begin{center}
      \resizebox{\hsize}{!}{\includegraphics[trim=200 360 0 130,clip]{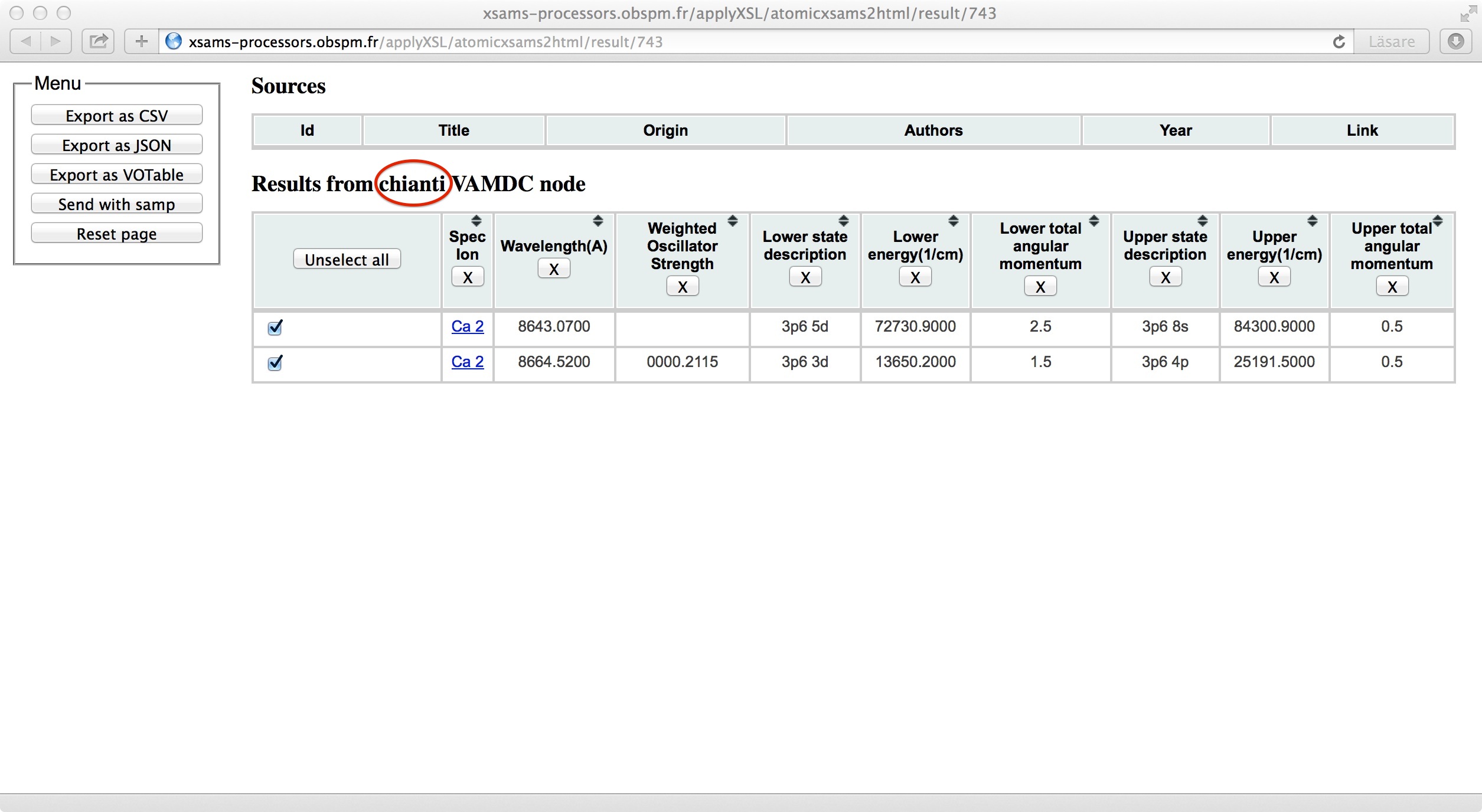}}
      \resizebox{\hsize}{!}{\includegraphics[trim=200 15 0 640,clip]{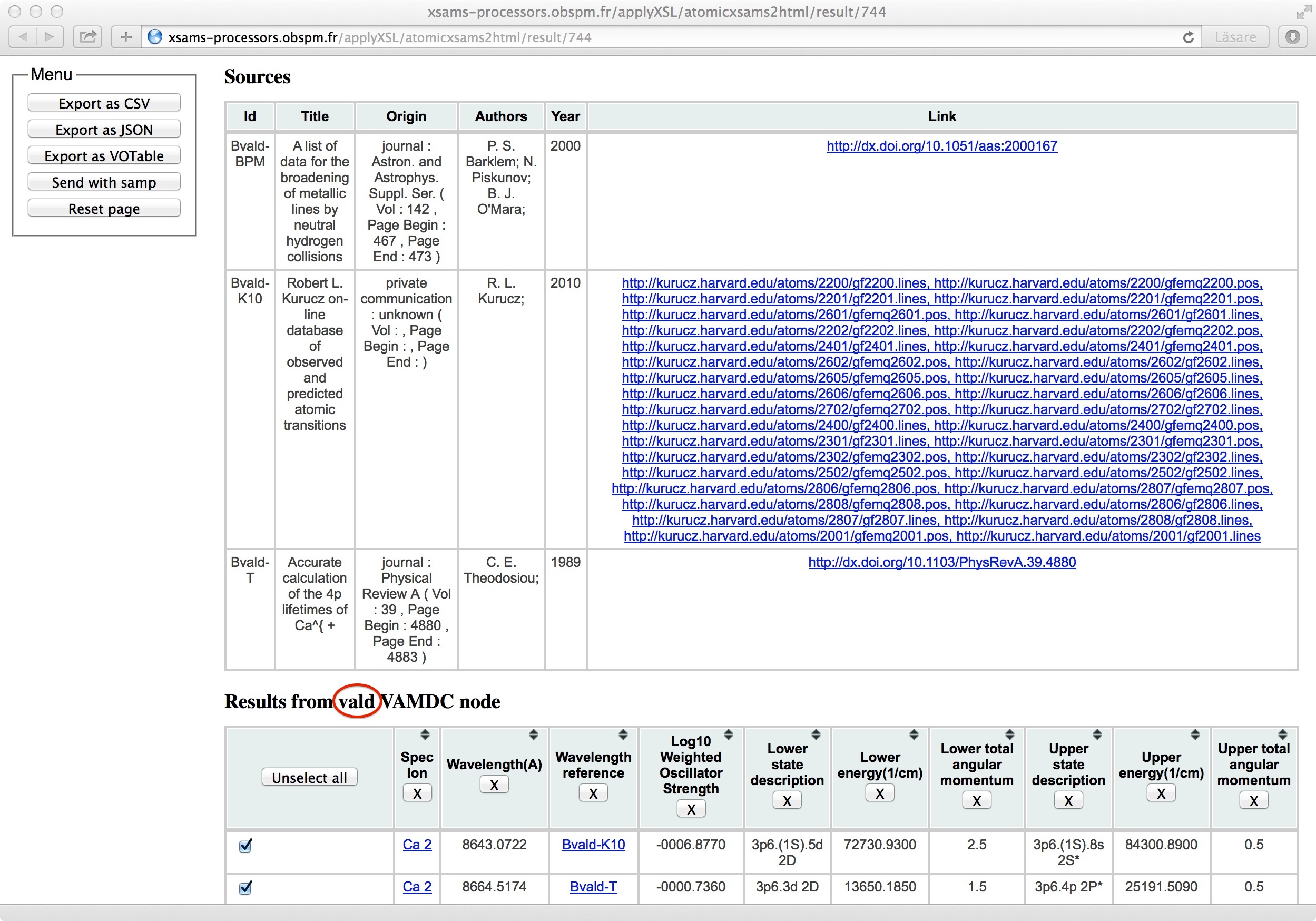}}
      \caption{Results of a query for radiative transition data of Ca$^+$ in the wavelength interval from 864 to 867~nm using the VAMDC graphical interface at \url{http://portal.vamdc.org/vamdc_portal/}. The data sources are given in a separate table.}
   \label{fig:VAMDC}
   \end{center}
\end{figure}

As an example, we describe a simple query for data for ionized calcium using the VAMDC portal. The web portal shows a list of all databases and dynamically highlights those which are available to answer the constructed query. It also provides a graphical interface, which allows one to select a species -- e.g. an atom with symbol Ca and ion charge 1, and the type of process, e.g. radiative transitions, and to specify further constraints. Setting the lower and upper wavelength limits to 864 and 867~nm and executing the query will result in a list of databases with an overview of the query results for each database (number of different species, states, and processes satisfying the constraints), as well as a datafile with the results for each database in a standard format.
The web portal can be used to display the results in tabular form. In the simple example, two transitions of Ca$^+$ are found in two databases (Chianti and VALD), including wavelengths, oscillator strengths, and state descriptions (see Fig.~\ref{fig:VAMDC}). In addition, the VALD database node returns detailed references to the data sources. 

\section{Gaia-ESO Public Spectroscopic Survey}

We illustrate the current status of atomic data for stellar spectrum modelling in the optical region with the example of the Gaia-ESO Public Spectroscopic Survey\footnote{\url{http://www.gaia-eso.eu/}} \cite{2012Msngr.147...25G}.
The goal of the Gaia-ESO survey is to provide a homogeneous overview of the distributions of motions and chemical abundances in the Milky Way. The survey is carried out as an ESO programme with more than 300 co-investigators, using the multi-object instrument FLAMES at the 8-m Very Large Telescope in Chile. About $10^5$ stars within the Milky Way, in the field or in stellar clusters, will be observed over five years (having started in 2012). Spectra are obtained in several wavelength regions, mostly covering 480 to 680~nm and 850 to 900~nm at different resolutions ($R=\lambda/\Delta\lambda=$47\,000 and $<$20\,000). These spectra are analysed to determine velocities, surface temperatures, surface gravities, and chemical abundances.

\begin{figure}[ht]
   \begin{center}
      \resizebox{\hsize}{!}{\includegraphics[trim=50 0 50 50,clip]{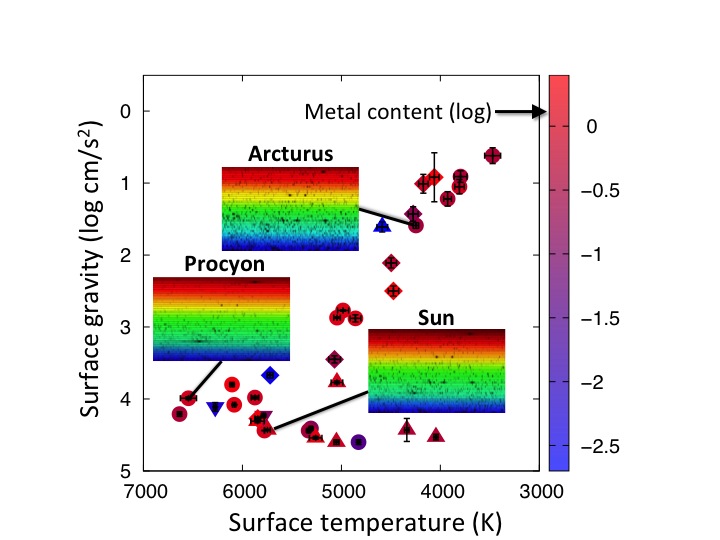}}
   \caption{The parameters of the Gaia FGK benchmark stars, illustrating the parameter ranges of typical Gaia-ESO targets. Optical spectra of three prototypical stars are shown as examples for the variation of absorption line patterns (Credit for inset images: N.A.Sharp, NOAO/AURA/NSF).}
   \label{fig:benchmarkstars}
   \end{center}
\end{figure}

All data are processed by several groups using different analysis codes. Early in the project it was decided that the spectrum analysis codes should use standardized input data as far as possible, such as model atmospheres, and atomic and molecular data. This approach allows one to rule out input data as sources of possible discrepancies between the results of different codes. Thus, we needed to define a standard line list, and to compile the best atomic data to be used by all groups. This task was assigned to a group of initially eight people within the Gaia-ESO collaboration. The first step was to identify spectral lines which allow accurate determination of stellar parameters and abundances of many chemical elements for F-, G-, and K-type stars.
Most of the Gaia-ESO targets belong to these stellar spectral types.
Figure~\ref{fig:benchmarkstars} illustrates the parameter ranges of typical Gaia-ESO targets by showing the parameters of the Gaia FGK benchmark stars \cite{2015arXiv150606095H,2014A&A...566A..98B,2014A&A...564A.133J}. The insets show the increasing number of absorption lines appearing in the optical spectra for decreasing surface temperatures. The Gaia-ESO spectra cover the green-to-red parts of the images, and a section in the near infrared.

\section{A line list for Gaia-ESO}
The approach for defining the Gaia-ESO line list was as follows. First, a list of 1341 preselected transitions for 35 elements (44 species comprising neutral and singly ionized atoms) was created\footnote{The species are H, Li, C, O, Na, Mg, Al, Si, Si$^+$, S, Ca, Ca$^+$, Sc, Sc$^+$, Ti, Ti$^+$, V, V$^+$, Cr, Cr$^+$, Mn, Fe, Fe$^+$, Co, Ni, Cu, Zn, Sr, Y, Y$^+$, Zr, Zr$^+$, Nb, Mo, Ru, Ba$^+$, La$^+$, Ce$^+$, Pr$^+$, Nd$^+$, Sm$^+$, Eu$^+$, Gd$^+$, Dy$^+$.}. 
Next, the best atomic data were selected for these lines from the literature.
We preferred the most precise laboratory measurements of $gf$-values (oscillator strengths), where available. This work led to a collaboration with the Laboratory Astrophysics group at Imperial College London, resulting in new data for neutral Fe lines \cite{2014MNRAS.441.3127R}. 
These data were supplemented by less precise laboratory $gf$-values and calculated data. We did not include or derive any astrophysical $gf$-values, as these would be dependent on the reference object and models used, and would not be applicable to all targets and analysis groups. Finally, simple quality flags for recommended use were assigned to each line (\emph{Yes}/\emph{Undecided}/\emph{No}), based on the quality of the selected data and evaluated with spectral syntheses for the Sun and Arcturus.

As an example, we describe the data selection for neutral Fe (the details for all species will be presented in a forthcoming publication).
The preselected line list includes 545 Fe lines.
For 42\% of these lines, 
precise laboratory measurements \cite{BKK,BK,BIPS,GESB79b,GESB82c,GESB82d,GESB86,BWL,2014MNRAS.441.3127R,2014ApJS..215...23D}, 
were available, and they were assigned the usage flag \emph{Yes}.
Older, less precise laboratory data \citeaffixed{MRW}{mainly by} 
were used for 33\% of the lines, 
which were assigned the usage flag \emph{Undecided}.
Finally, the semi-empirical calculations by \citeasnoun{K07} 
were assigned to the remaining 25\% of the lines 
together with the usage flag \emph{No}.

\begin{figure}[ht]
   \begin{center}
   \resizebox{\hsize}{!}{\includegraphics[trim=100 67  96 68,clip]{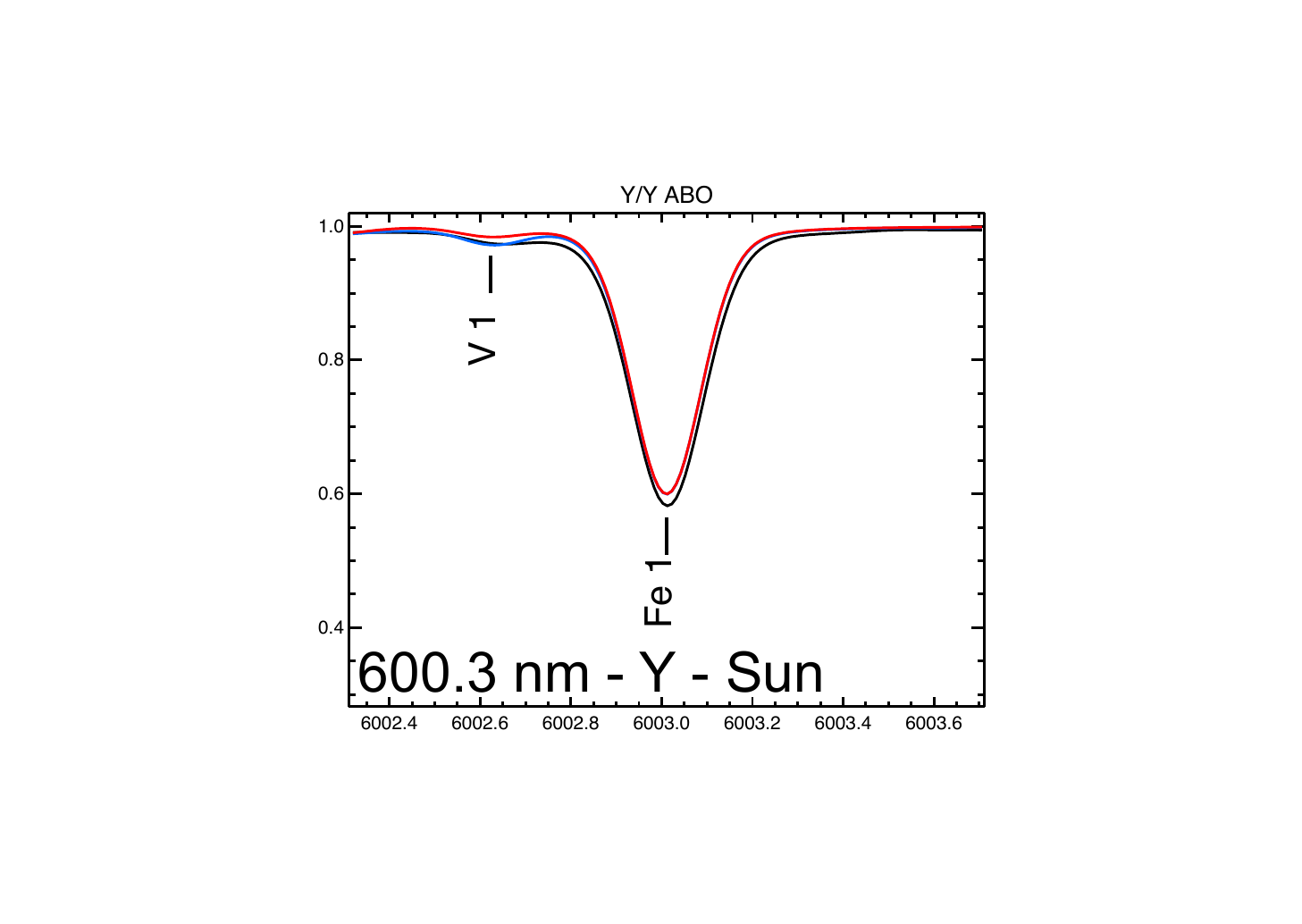}\includegraphics[trim=100 67  96 68,clip]{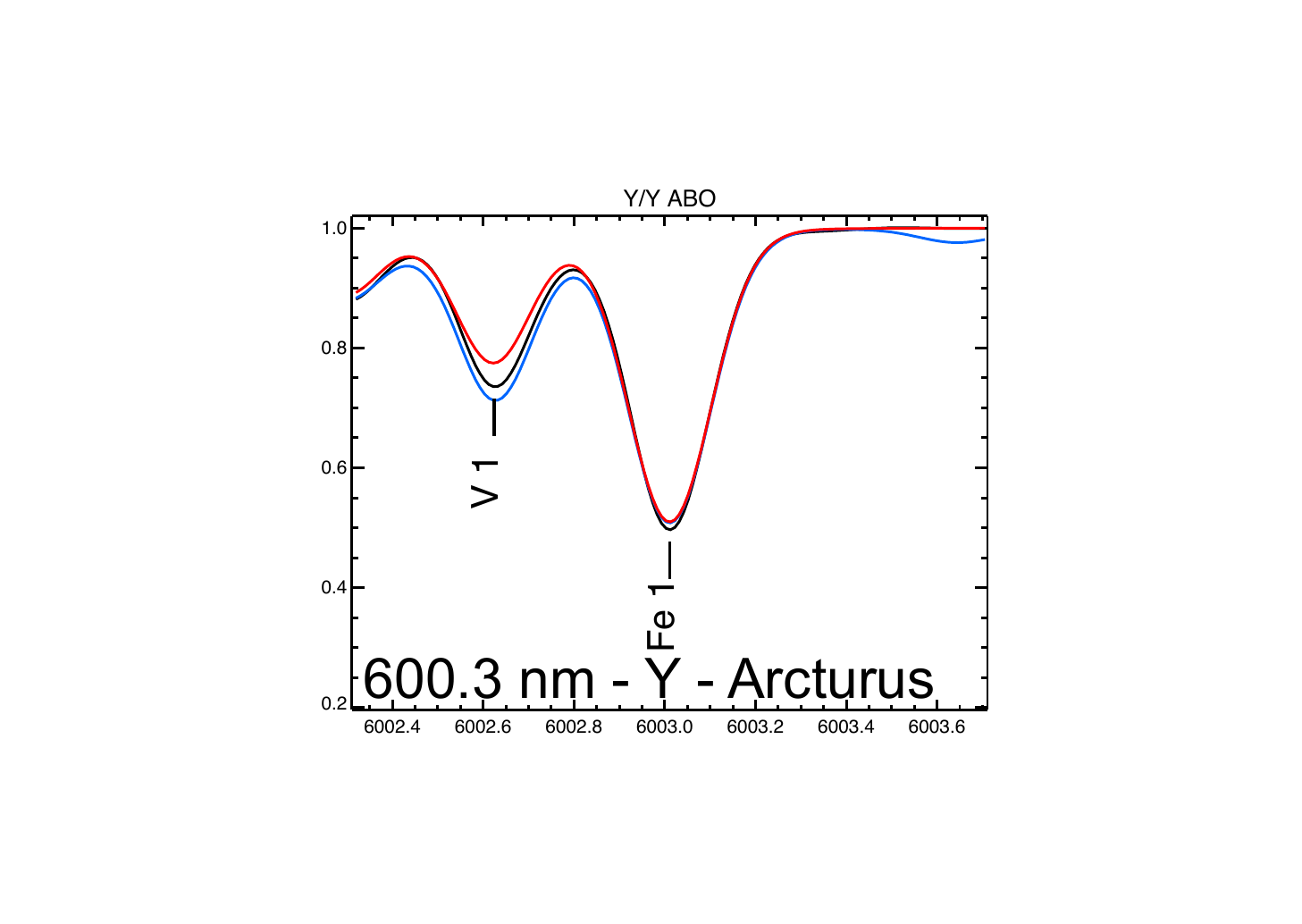}}
   \resizebox{\hsize}{!}{\includegraphics[trim=100 69  96 68,clip]{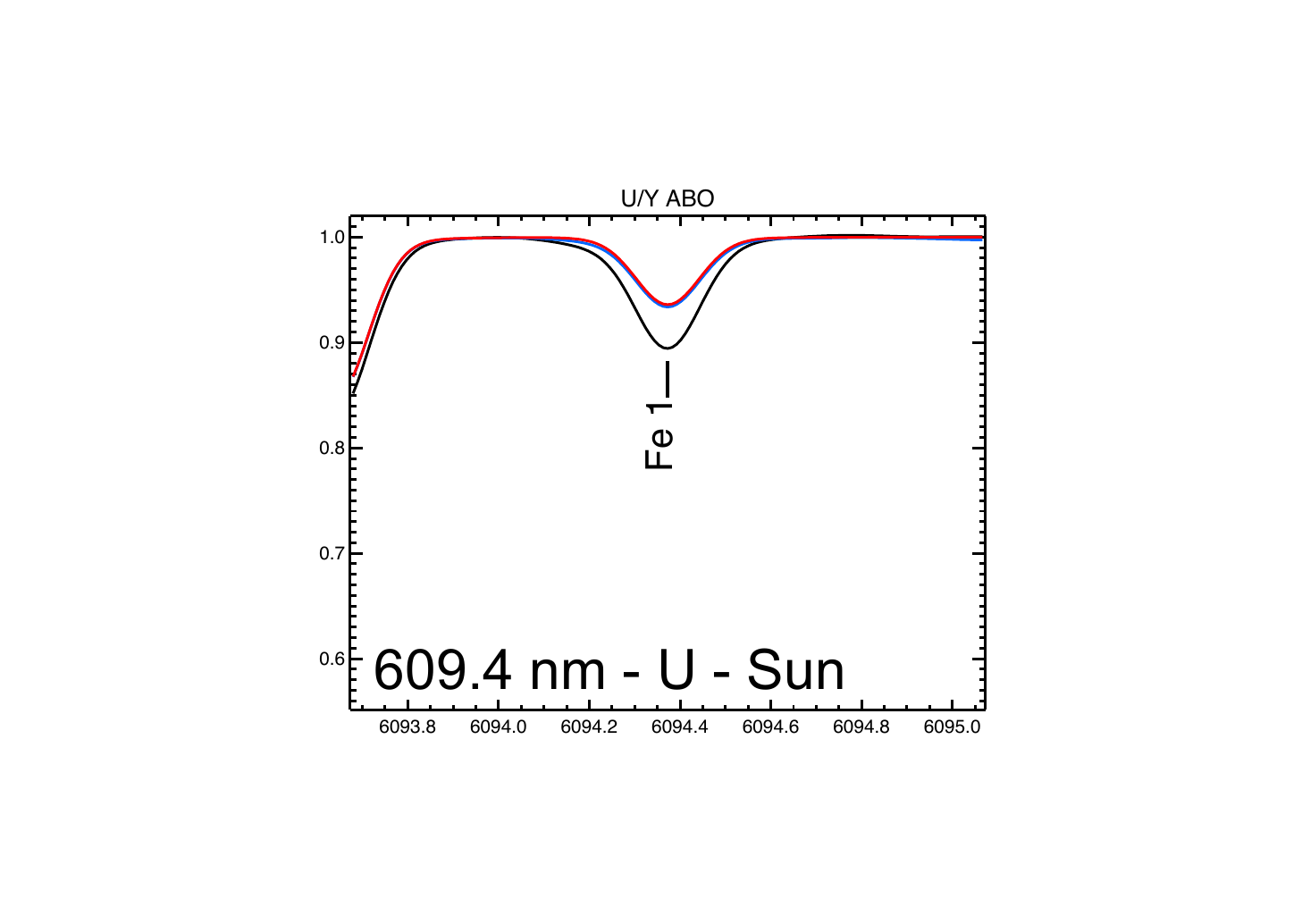}\includegraphics[trim=100 69  96 66,clip]{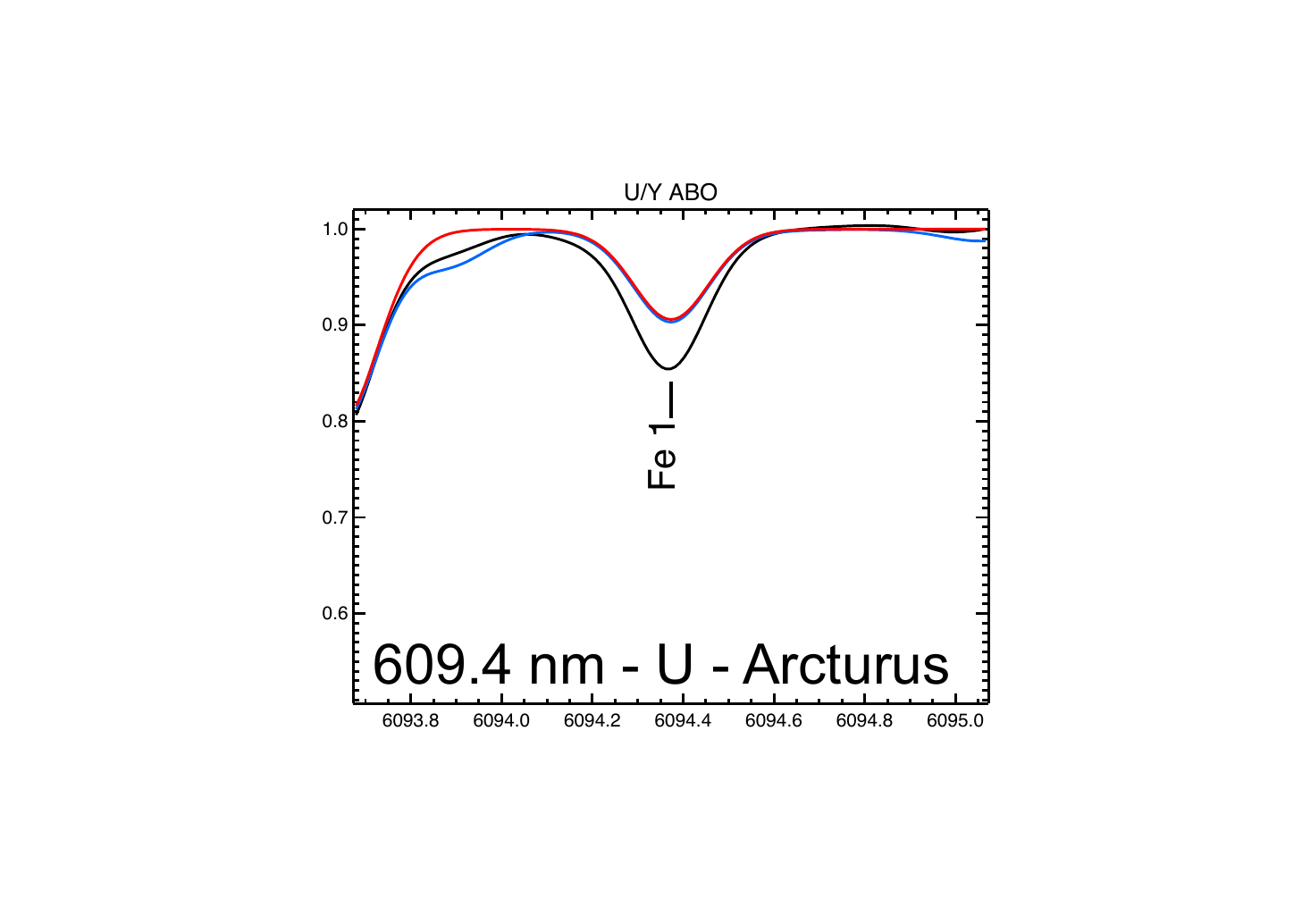}}
   \resizebox{\hsize}{!}{\includegraphics[trim=100 69 100 66,clip]{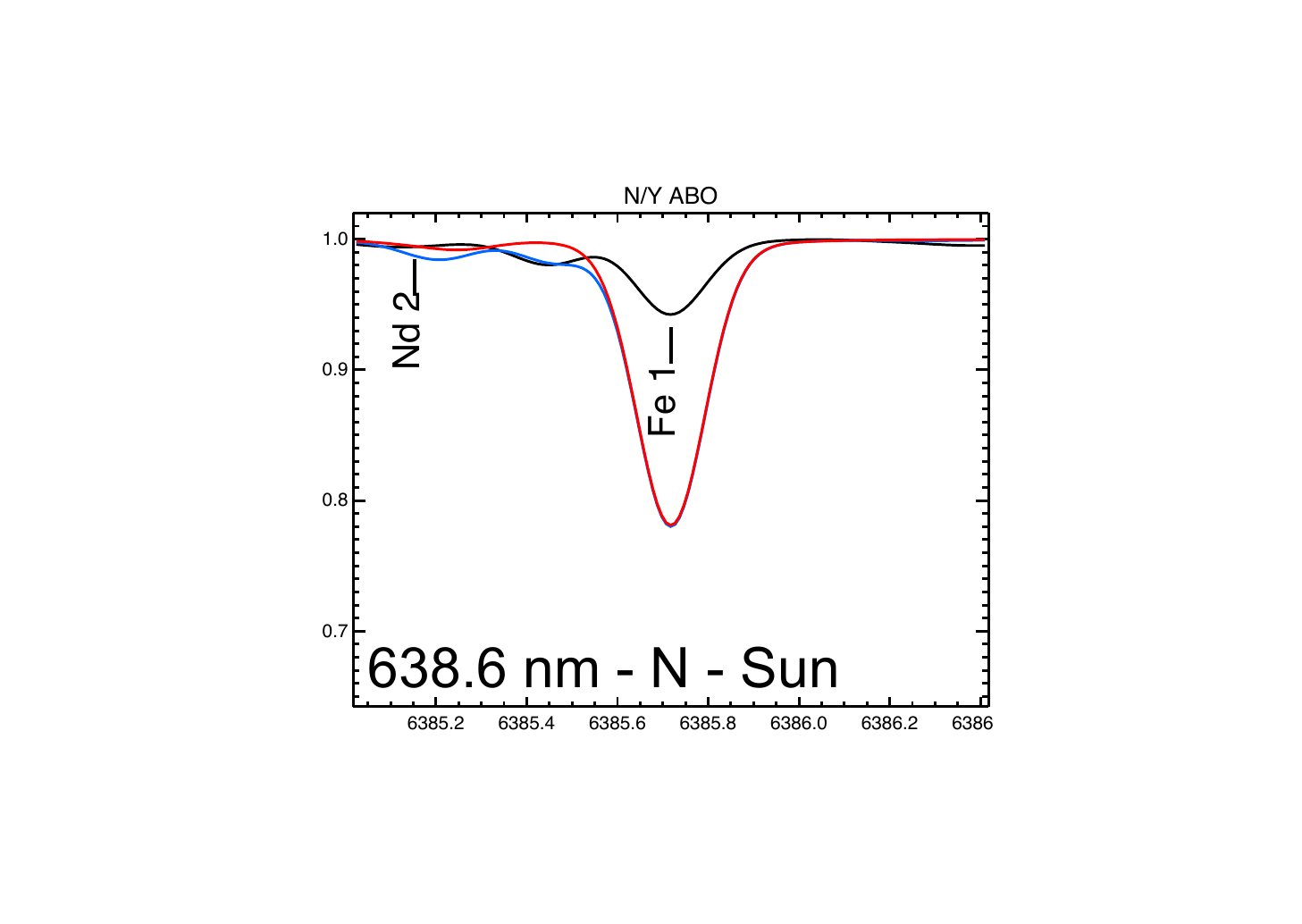}\includegraphics[trim= 90 69 100 66,clip]{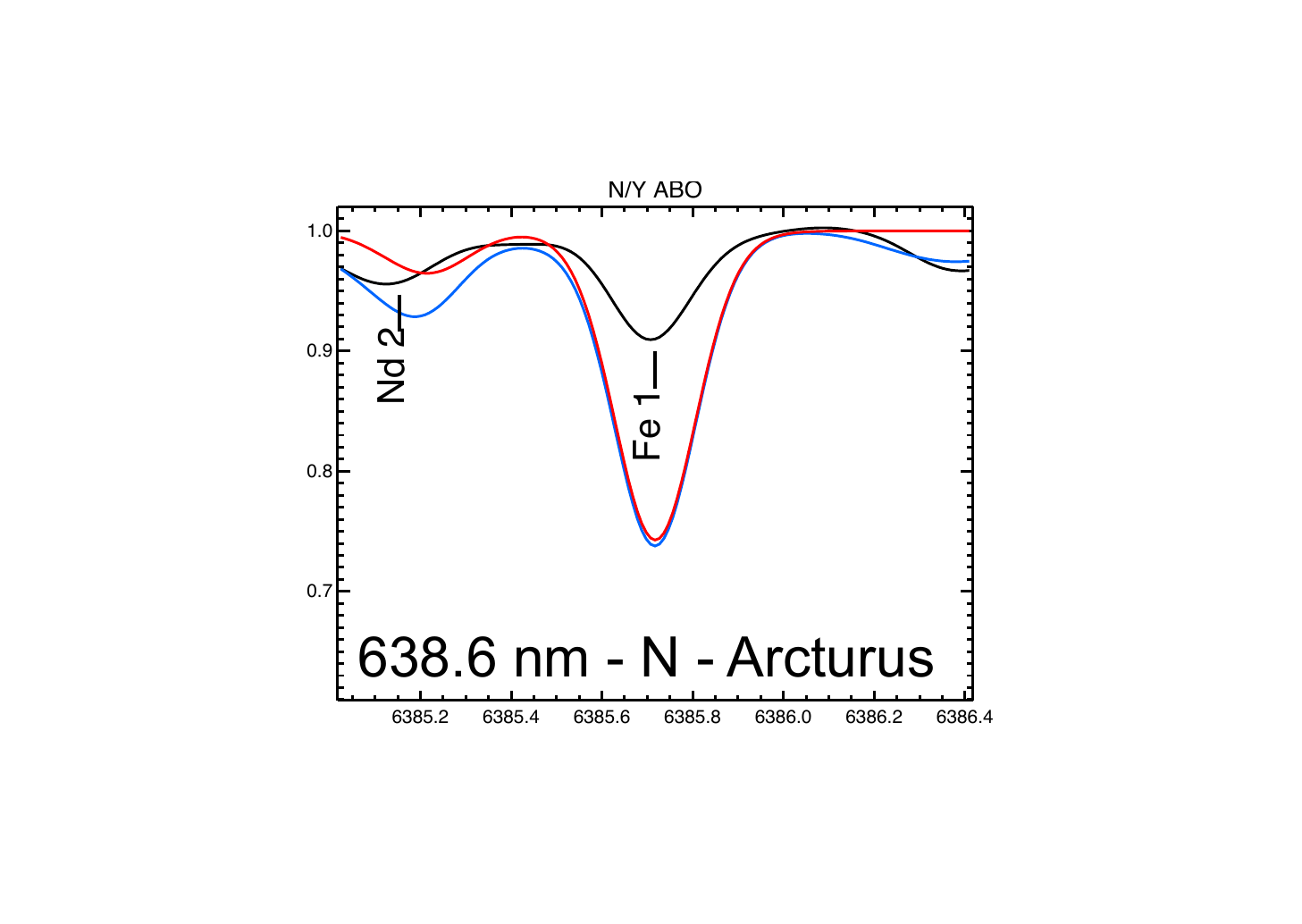}}
   \caption{Comparison of observed and synthetic spectra around three Fe lines with different flags (Y for \emph{Yes}, U for \emph{Undecided}, N for \emph{No}) for the Sun (left) and Arcturus (right). Black lines: observations, red lines: calculations including preselected spectral lines only, blue lines: calculations including blends.}
   \label{fig:Fe1}
   \end{center}
\end{figure}

Figure~\ref{fig:Fe1} shows example spectra for three Fe lines with different flags for recommended usage for the Sun and Arcturus. Observed spectra are from \citeasnoun{1984sfat.book.....K} and \citeasnoun{2000vnia.book.....H}, 
respectively and were convolved to $R=$47\,000.
For the calculated spectra MARCS atmospheric models \cite{Gust:08} 
and solar abundances by \citeasnoun{2007SSRv..130..105G} 
were used.
In the case of Arcturus, the model parameters are $T_{\rm eff}$=4247~K and $\log g$(cm s$^{-2}$)=1.59, and the abundances were reduced by $-$0.54~dex, with 0.2~dex enhancement for $\alpha$-process elements.
Observed and synthetic spectra agree for both stars in the case of the line with the \emph{Yes} flag, while they disagree to different degrees for the lines with \emph{Undecided} and \emph{No} flags.
%
The astrophysical performance of lines with different flags was investigated through
abundance determinations for each individual Fe line using the observed spectrum of the Sun. The lines with a \emph{Yes} flag showed a somewhat smaller dispersion around the mean than the lines with an \emph{Undecided} flag. Abundances for lines with a \emph{No} flag deviated from the mean by up to 1.5~dex.

Even if the analysis focuses on wavelength regions around the recommended list of preselected lines, these data are not sufficient for a thorough analysis. We need complete information (as far as possible) on all transitions visible in the observed wavelength ranges, in order to identify blends of the preselected lines, as ``background'' for synthetic spectrum calculations, and for an evaluation of the quality of spectrum processing (e.g. continuum normalization). Therefore, the preselected lines were complemented with data for $\sim$72\,000 atomic lines extracted from the VALD database with default configuration and stellar parameters corresponding to those of the target stars, as well as data for 27 molecular species.
Priority was given to molecules which contribute significantly to the absorption in the spectra of G or K-type stars. For TiO, ZrO, SiH, CaH, VO, and FeH (and their isotopologues), the best line lists available in the literature were used. For CH, C$_2$, NH, OH, and MgH, improved line lists were computed using recent laboratory measurements. A detailed explanation of the procedure for the case of CH can be found in \citeasnoun{2014A&A...571A..47M}. 

\begin{figure}[ht]
   \begin{center}
      \resizebox{\hsize}{!}{\includegraphics{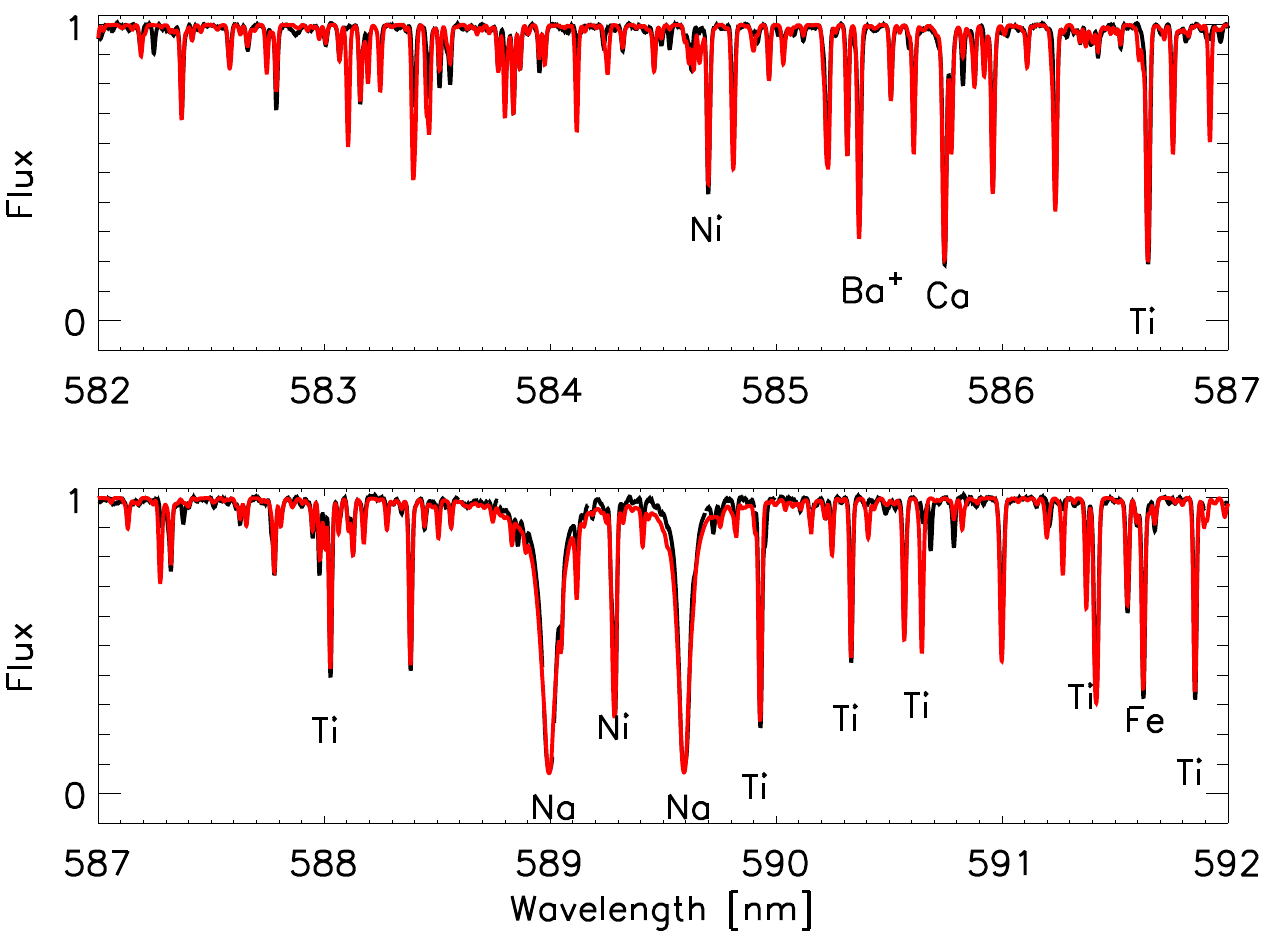}}
      \caption{Observed (black) and calculated (red) spectra for Arcturus around the Na doublet lines at 589~nm. The calculations include the full Gaia-ESO line list.}
      \label{fig:master}
   \end{center}
\end{figure}

Figure~\ref{fig:master} shows observed \cite{2000vnia.book.....H} 
and calculated spectra for Arcturus, convolved to $R=$47\,000 in an interval of 10~nm near the Na doublet lines at 589~nm, using the full line list. Most of the observed features are reproduced by the calculations, but in several places the calculated lines are too weak or completely missing, indicating a lack of atomic data.

\section{Summary and Outlook}
Accurate laboratory data in the optical and IR wavelength regions are needed for the analysis of stellar spectra from ongoing or planned large-scale surveys.
The Gaia-ESO collaboration provides a list of recommended lines for the analysis of FGK stars. The Gaia-ESO line list is regularly updated, resulting in a new version about once a year.
The tests of the performance of the preselected lines should be extended to all of the Gaia FGK benchmark stars. Work in this direction has started within the Gaia-ESO collaboration.
It is worth noting that numerous lines in the spectra of FGK stars are still unidentified.
This problem can be remedied either by analysis of laboratory spectra, or analysis of carefully selected stellar spectra \citeaffixed[for a novel approach to energy-level determinations for Fe lines]{2015ApJS..216....1P}{see e.g.}. 

\ack
We acknowledge the contributions of H.~J\"onsson, P.~de~Laverny, E.~Maiorca, N.~Ryde, and J.~Sobeck to the Gaia-ESO line list work.
UH acknowledges support from the Swedish National Space Board (Rymdstyrelsen).
JCP and MR thank STFC of the UK for support of the laboratory astrophysics programme at Imperial College.

\section*{References}
\bibliographystyle{jphysicsB}
\bibliography{heiter}

\end{document}